# A Via-less Fully Screen-Printed Reconfigurable Intelligent Surface for 5G Millimeter Wave Communication

Yiming Yang[(1)], Ruiqi Wang[(1)], Mohammad Vaseem[(1)], Behrooz Makki[(2)], and Atif Shamim[(1)]

(1) Computer, Electrical, and Mathematical Sciences and Engineering Division
King Abdullah University of Science and Technology, Thuwal, KSA
(yiming.yang@kaust.edu.sa, ruiqi.wang.1@kaust.edu.sa, mohammad.vaseem@kaust.edu.sa,
atif.shamim@kaust.edu.sa)
(2) Ericsson Research, 417 56 Gothenburg, Sweden, https://www.ericsson.com/

*Abstract*—In this paper, we propose a via-less fully screen-printed reconfigurable intelligent surface which can establish a second line-of-sight communication from 23.5GHz to 29.5GHz. By serially connecting the H shaped resonator along the H field of the incident wave, we minimize the effect of the biasing lines and make a via-less design, which reduces the fabrication difficulty and cost. The unit-cell simulation of the array with screen-printed $VO_2$ switches shows a 215° to 160° phase shift difference between the ON and OFF states within bandwidth. During the field testing of the ideal arrays, we verify that the array can redirect the 45° incident wave to 0° reflection with a signal enhancement of at least 10 dB as compared to the array which has all unit cells in the OFF condition.

## I. Introduction

The 5G and beyond millimeter wave communication is gaining more and more attraction thanks to its high bandwidth and low latency [1]. However, as the wavelength decreases to millimeters, the signal path has an increased loss and can be easily blocked by obstacles in direct-link communications. To overcome this problem, the reconfigurable intelligent surface (RIS) is proposed to fix the coverage holes by controlling the surface phase distribution and redirecting the reflected beam to the receiver [1].

A RIS consists of an array of switchable elements, and its performance enhances as the array size increases [1]. Usually, the implementations of reconfigurability involve active switches [2] such as PIN diodes and MEMS, or tunable devices such as varactors, etc. They can indeed provide low loss and good reconfigurability in phase, but the cost increases significantly when the array size is large. What's more, most RIS designs require at least one layer of resonators and another layer of biasing lines, which introduce vias and increase loss, fabrication difficulty, and cost.

To have a large RIS array with low cost and ease of fabrication, we propose a via-less RIS operating from 23.5GHz to 29.5GHz with fully screen-printed $VO_2$ switches [3] as well as silver (Ag) paste conductors. The unit cell of the array combines the resonator as well as the serial biasing network, which only requires one layer of the screen-printing process [4] without vias.

## II. Unit Cell Design and Simulation

The unit cell design is shown in Fig. 1. The RIS consists of three layers, which are the ground plane, the substrate, and the resonator. The resonator is H-shaped with a wide main dipole branch and narrow biasing branches. The main branch is in parallel with the polarization of the incidence. In the middle of the dipole main branch, as shown in the green sections in Fig. 1, the $VO_2$ switches are screen-printed to control the states of the resonators. The switch-on resistance is 4 Ω and the switch-off resistance is 1000 Ω. Along the H field direction, the resonators are serially connected and each column of resonators can be controlled with only one biasing line.

For ease of printing and testing, we place the resonators on a 50 um PEN sheet and then attach the sheet to the substrate. The substrate is 0.8mm AF32 eco glass with a permittivity of 5.1 and a loss tangent of 0.0086 at 28.5GHz. The ground plane and the conductor part of the resonator are modeled with 50 um Ag paste conductor and the conductivity is $7 \times 10^6$ S/m.

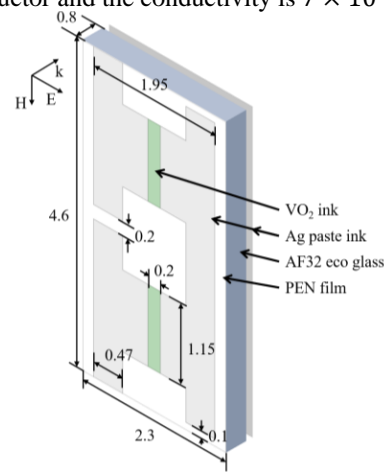

Figure 1. Unit cell design (dimensions are in millimeters)

As the unit cell periodicity decreases, the reflection magnitude and bandwidth increase [5]. Since the screen-printed materials have a higher conductor loss and switch resistance loss than copper and active switches, the unit cell should be reduced compared with the traditional PCB-based

RIS. The periodicity of the resonator is 2.3mm, which is 36.1% of the guided wavelength at 29.5 GHz.

The simulated phase shift and magnitude of the unit cell with periodic boundary conditions and Floquet port excitation are plotted in Fig. 2(a). As frequency increases from 23.5 GHz to 29.5 GHz, the phase shift difference between the switch-off and the switch-on states decreases from 215° to 160°. The switch-off magnitude decreases from 0.94 to 0.88, and the switch-on magnitude increases from 0.57 to 0.74. The loss of the switch-on state is contributed by the $VO_2$ switch-on resistance.

To see the effect of the biasing lines, the surface current distributions of both ON and OFF states are plotted in Fig. 2(b) and (c). At letter A in Fig. 2(b), the biasing line connects the two H-shaped resonators. For both ON and OFF states, the connection segments have the minimum resonant current and thus they have little effect on the resonator.

The phase shift difference between the two states can also be explained with the surface current distributions. When the switch is ON, the $VO_2$ labeled in letter B in Fig. 2(b) connects the two segments of the resonator and lets current flow through. Thus, the resonator has a longer current path and a lower resonant frequency. When the switch is OFF, the switch isolates the two segments and the resonant path becomes much shorter. Thus, the resonant frequency increases.

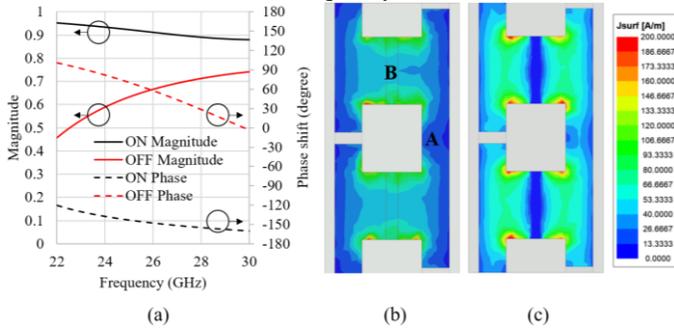

Figure 2. Simulated results of the unit cell. (a) reflection magnitude and phase. (b) ON state and (c) OFF state surface current distribution at 27.5 GHz.

### III. ARRAY FABRICATION AND MEASUREMENT

As an initial proof of concept and to estimate the performance limit of the final array with printed $VO_2$ switches, we print a 20×20 ideal array of resonators on PEN sheets with commercial Ag paste only, which is shown in Fig. 3(a). For different array configurations, we print different combinations of ON and OFF resonator columns. The ON resonator column has the Ag paste ink printed in the letter B region in and at the OFF column, we leave the $VO_2$ switch unprinted as a slot. As Fig. 3(b) shows, we attach the printed RIS film to the glass substrate, which is backed by copper tapes as the ground plane.

As Fig. 3(b) shows, the ON array pattern is demonstrated with black and red blocks, where the ON columns are labeled in black and the OFF columns are in red. The column patterns are determined using the following equation [2].

$$\Delta\varphi = \frac{2\pi}{\lambda}\left(\overline{r_{tx}} - \overline{r_{column}} - \overline{r_{reflect}}\right)\cdot\overline{r_{column}} \quad (1)$$

In (1), $\lambda$ is the free space wavelength at 27.5GHz. $\overline{r_{tx}}$ is the position of the Tx antenna. $\overline{r_{column}}$ is the center position of the column to be configured. $\overline{r_{reflect}}$ is the normalized reflection direction vector. When $\Delta\varphi$ is closer to the switch-off instead of the switch-on phase shift at 27.5 GHz in Fig. 2(a), the column uses an OFF pattern, and vice versa.

The measurement setup is shown in Fig. 3(b). A pair of 20 dBi standard horn antennas are placed in front of the RIS array. The transmitting (Tx) antenna is placed 20 cm away from the RIS with a 45° incident angle. The receiving (Rx) antenna is placed in the normal direction of the RIS at 20 cm away. The Tx antenna is connected to port 1 of a VNA and the Rx antenna is connected to port 2. The S21 is measured as the relative received power after the Tx power is reflected by the RIS in the normal direction.

The measurement result is shown in Fig. 3(c). We can see that from 23.5 GHz to 29.5 GHz, the array has at least 10 dB signal enhancement when switched from all-column-OFF patterns to ON patterns. The ripples of the ON RIS are contributed by the unit cells near the end of the columns. Compared with the unit cells at the array center, the distance from those at the end of columns to the Rx and Tx antennas is larger than that of the center unit cells, which introduces additional phase shift error.

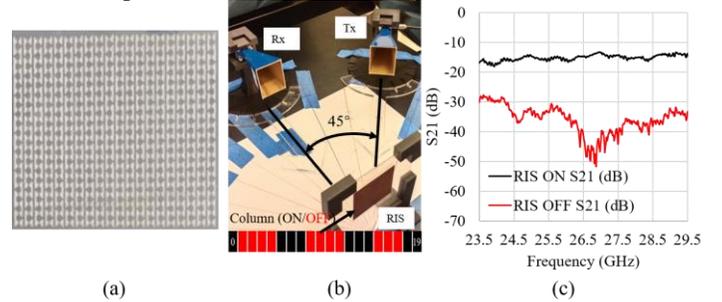

Figure 3. The ideal array (a) sample, (b) measurement setup, and (c) results.

### IV. CONCLUSION

In conclusion, we have designed via-less fully screen-printed RIS unit cells with a single top layer combining resonators and serial biasing lines. We have also demonstrated an ideal 20x20 RIS array, which can direct the reflection beam from a 45° incident angle to a 0° reflection angle with at least 10 dB signal enhancement from 23.5 GHz to 29.5 GHz. The array design is suitable for large-scale screen-printing processes and is easy to increase the size while maintaining simplistic serial biasing lines for each column of resonators.